\documentclass[amsmath,amssymb,prb,reprint,superscriptaddress]{revtex4-2}

\usepackage{physics}
\usepackage{amsmath}
\usepackage{xparse}
\usepackage{graphicx}
\usepackage{mathtools}

\newcommand{\settitle}{
    \title{General Strategy for Large Nernst Coefficient}

    \author{Junya Endo}
    \affiliation{Department of Physics, University of Tokyo, Bunkyo, Tokyo, Japan}

    \author{Hiroyasu Matsuura}
    \affiliation{National Institute of Advanced Industrial Science and Technology (AIST), Tsukuba, Ibaraki, Japan}

    \author{Manfred Sigrist}
    \affiliation{Department of Physics, ETH Zurich, 8093 Zurich, Switzerland}

    \author{Masao Ogata}
    \affiliation{Department of Physics, University of Tokyo, Bunkyo, Tokyo, Japan}
    \affiliation{National Institute of Advanced Industrial Science and Technology (AIST), Tsukuba, Ibaraki, Japan}

    \date{\today}
}

\newcommand{\subfigimg}[3][,]{%
    \setbox1=\hbox{\includegraphics[#1]{#3}}
    \leavevmode\rlap{\usebox1}
    \rlap{\hspace*{-0.02\linewidth}\raisebox{\dimexpr\ht1-1\baselineskip\relax}{\normalsize\textbf{(#2)}}}
    \phantom{\usebox1}
}

\newcounter{num}
\newcommand{\Rnum}[1]{\setcounter{num}{#1} \Roman{num}}

\newcommand{\ep}{\varepsilon}
\newcommand{\vk}{\vb*{k}}
\newcommand{\vq}{\vb*{q}}
\newcommand{\vrr}{\vb*{r}}

\newcommand{\Boltzmann}{k_{\mathrm{B}}}
\newcommand{\Temperature}{T}
\newcommand{\KBT}{\Boltzmann \Temperature}
\newcommand{\InvKBT}{\beta}
\newcommand{\Volume}{V}

\newcommand{\Seebeck}{S}
\newcommand{\Nernst}{N}
\newcommand{\CondRatio}{R}
\newcommand{\CondTh}{\kappa}
\newcommand{\CondThMat}{\overline{\kappa}}
\newcommand{\zL}{z}

\newcommand{\CurrentEle}{\vb*{J}^{1}}
\newcommand{\CurrentTh}{\vb*{J}^{2}}
\NewDocumentCommand{\CurrentDensEle}{ o }{\vb*{j}^{1}\IfValueT{#1}{_{#1}}}

\NewDocumentCommand{\FieldEle}{ o }{\IfValueTF{#1}{E_{#1}}{\vb*{E}}}
\NewDocumentCommand{\GradientTemperature}{ o }{\IfValueTF{#1}{\pqty{\grad T}_{#1}}{\grad T}}
\newcommand{\FieldTemperature}{\frac{\GradientTemperature}{\Temperature}}

\newcommand{\Chemical}{\mu}
\newcommand{\EleCharge}{e}
\newcommand{\MassEle}{m}
\NewDocumentCommand{\Fermi}{ o }{f\IfValueT{#1}{\pqty{#1}}}
\NewDocumentCommand{\FermiDeriv}{ o }{\Fermi'\IfValueT{#1}{\pqty{#1}}}
\NewDocumentCommand{\Velocity}{ o m o }{\IfValueTF{#1}{v}{\vb*{v}}_{\IfValueT{#3}{#3,}#2\IfValueT{#1}{,#1}}}
\NewDocumentCommand{\Bloch}{ m o }{u_{
            \IfValueT{#2}{#2,}#1}}

\newcommand{\LMat}[1][]{\overline{L}^{#1}}
\newcommand{\LDiag}[1][]{L^{#1}_{xx}}
\newcommand{\LNDiag}[1][]{L^{#1}_{xy}}
\newcommand{\LElem}[2][]{L^{#1}_{#2}}

\NewDocumentCommand{\Sp}{ m o }{\sigma_{#1}\IfValueT{#2}{\pqty{#2}}}
\NewDocumentCommand{\SpL}{ o }{\sigma_{xx}\IfValueT{#1}{\pqty{#1}}}
\NewDocumentCommand{\SpT}{ o }{\sigma_{xy}\IfValueT{#1}{\pqty{#1}}}
\NewDocumentCommand{\Mean}{ m m }{\left\langle #2 \right\rangle_{#1}}
\NewDocumentCommand{\MeanL}{ m }{\left\langle #1 \right\rangle_{xx}}
\NewDocumentCommand{\MeanT}{ m }{\left\langle #1 \right\rangle_{xy}}

\newcommand{\Int}[2]{\int #2 ~ \dd{#1}}
\newcommand{\IntReg}[4]{\int_{#1}^{#2} #4 ~ \dd{#3}}
\newcommand{\OrderLandau}{\mathcal{O}}
\newcommand{\exps}[1]{\mathrm{e}^{#1}}
\newcommand{\PolyLog}[2]{\mathrm{Li}_{#1}\pqty{#2}}
\newcommand{\Div}{\grad \cdot}

\newcommand{\Hamiltonian}{\mathcal{H}}
\newcommand{\HamiltonianDensity}{h}
\NewDocumentCommand{\OpC}{ o }{\IfValueTF{#1}{\vb*{\hat{c}}_{#1}}{\vb*{\hat{c}}}}
\NewDocumentCommand{\OpCD}{ o }{\IfValueTF{#1}{\vb*{\hat{c}}^{\dagger}_{#1}}{\vb*{\hat{c}}^{\dagger}}}
\newcommand{\OpCurrentEle}{\vb*{\hat{J}}^{1}}
\newcommand{\OpCurrentTh}{\vb*{\hat{J}}^{2}}
\newcommand{\OpCurrentHeat}{\vb*{\hat{J}}^{2}}
\newcommand{\OpNum}{\vb*{\hat{N}}}
\newcommand{\OpDens}{\hat{\rho}}
\NewDocumentCommand{\TauOrder}{o}{\mathrm{T}_{\tau} \IfValueT{#1}{\bqty{#1}}}
\newcommand{\mean}[1]{\left\langle #1 \right\rangle}
\NewDocumentCommand{\Correlation}{ o o o o }{\Phi\IfValueT{#1}{^{#1}}\IfValueT{#2}{_{#2}}\IfValueT{#3}{\pqty{#3 \IfValueT{#4}{, #4}}}}

\newcommand{\ien}{i\varepsilon_{n}}
\newcommand{\iwlam}{i\omega_{\lambda}}
\NewDocumentCommand{\GreT}{ o o }{\mathcal{G}\IfValueT{#1}{_{#1}}\IfValueT{#2}{\pqty{#2}}}
\NewDocumentCommand{\GreR}{ o o }{G^{R}\IfValueT{#1}{_{#1}}\IfValueT{#2}{\pqty{#2}}}
\NewDocumentCommand{\GreA}{ o o }{G^{A}\IfValueT{#1}{_{#1}}\IfValueT{#2}{\pqty{#2}}}

\begin{document}

\settitle

\begin{abstract}
    We propose a general strategy for enhancing the anomalous Nernst coefficient based on the Sommerfeld-Bethe relation. This approach provides a systematic framework for understanding the small anomalous Nernst coefficients typically observed in ferromagnets and identifies conditions under which substantial enhancements can be realized. We further introduce simplified models that exhibit large Nernst coefficients as offering illustrative examples.
\end{abstract}

\maketitle

\section{Introduction}
\label{section:introduction}
The anomalous Hall effect (AHE) is a cornerstone phenomenon in condensed matter physics; in ferromagnetic materials, an electric current generates a transverse voltage in the absence of an external magnetic field. First discovered by Edwin Hall in 1881~\cite{Hall1881}, following his discovery of the ordinary Hall effect in 1879~\cite{Hall1879}, the AHE has been instrumental in exploring the intrinsic properties of magnetic materials and has profound implications for spintronics and magnetoelectronics.

Initially, the AHE was considered to result from the influence of internal magnetization acting like an effective magnetic field on charge carriers~\cite{Pugh1953}. However, this classical interpretation was insufficient to explain the complexities observed in experiments. The pioneering work by Karplus and Luttinger in 1954~\cite{Karplus1954} introduced a quantum mechanical perspective, attributing the AHE to the intrinsic spin-orbit coupling (SOC) within the band structure of ferromagnets. They proposed that the anomalous velocity arising from SOC leads to a transverse deflection of charge carriers, laying the foundation for the intrinsic mechanism of the AHE.

Parallel to the developments in electric transport, the anomalous Nernst effect (ANE) has attracted significant attention. The ANE is the thermoelectric counterpart of the AHE; a temperature gradient induces a transverse electric field in magnetic materials. As in the AHE, the ANE arises from both intrinsic and extrinsic mechanisms linked to SOC and magnetic order. The intrinsic contribution to the ANE is also associated with the Berry curvature, reflecting the fundamental symmetry and topology of the electronic structure~\cite{Xiao2006}. In addition, as the extrinsic origin, it is also found that ANE is enhanced in the film-substrate system~\cite{Zhou2024,Matsuura2024}.

The ANE has conventionally been considered to scale with a material's magnetization\cite{Guin2019}. However, recent investigations have challenged this assumption, showing that the relationship between magnetization and ANC is more complex than previously thought. Recent studies demonstrate that certain topological magnets can exhibit substantial ANC values even with small or negligible net magnetization. A notable example is observed in the chiral antiferromagnets Mn$_{3}$Sn and Mn$_{3}$Ge, which were reported to display pronounced AHE~\cite{Nakatsuji2015,Kiyohara2016}. Subsequent to this discovery, a significant ANE was also demonstrated in these compounds. The observed anomalous Nernst coefficients (ANC) attain values of approximately $0.35~\mu \mathrm{V/K}$ at room temperature in Mn$_{3}$Sn~\cite{Ikhlas2017} and similar values in Mn$_{3}$Ge~\cite{Wuttke2019,Xu2020}, which are comparable to those observed in ferromagnetic metals. These phenomena are attributed to the Berry curvature near the Weyl points which appeared due to the symmetry breaking. In another instance, the ferromagnetic Heusler compound Co$_{2}$MnGa exhibited a significant enhancement of the ANC driven by a large Berry curvature near the Fermi level, as observed at room temperature~\cite{Guin2019,Nakatsuji2022}.

The Berry curvature enhances the anomalous Nernst conductivity through the same topological mechanism that governs the anomalous Hall conductivity. However, enhancing the ANC presents additional complexity. The Nernst coefficient is defined under conditions without electric currents, and one cannot conclude that it becomes large solely due to high anomalous Nernst conductivity. Nevertheless, the ANC remains a critical indicator for the efficiency of thermoelectric conversion, particularly in transverse thermoelectric devices. Understanding the mechanisms contributing to its enhancement is therefore essential for developing advanced thermoelectric materials.

In this paper, we investigate general conditions for the anomalous Nernst coefficient. For longitudinal thermoelectric effects, Mahan and Sofo proposed a general strategy for achieving large $\zL T$ through the diffusive Seebeck effect \cite{Mahan1996}. Analogously, a general strategy for large ANE is anticipated. We therefore examine general conditions for large ANE based on the diffusive Seebeck effect, following the Mahan-Sofo approach.

The paper is structured as follows. Section \Rnum{2} provides a rigorous derivation of the Sommerfeld-Bethe relation for both longitudinal and transverse response functions, establishing our theoretical framework. In Section \Rnum{3}, we employ this framework to develop a Mahan-Sofo-type optimization criterion that directly relates the spectral shapes of $\sigma_{xx}$ and $\sigma_{xy}$ to the achievable magnitude of the anomalous Nernst coefficient. Section \Rnum{4} examines this criterion through three representative models—conventional ferromagnets, a linear-band metal, and a Rashba-Zeeman semiconductor—identifying specific pathways through which $\abs{N}$ can approach the benchmark value $\Boltzmann/\abs{\EleCharge}$. Section \Rnum{5} concludes with a summary of key findings and proposes design principles for developing materials with exceptionally large transverse thermoelectric responses.

\section{Sommerfeld-Bethe relation}
\label{section:SB}
In this paper, we examine the linear-response regime of thermoelectric effects. We refer to the electric current and thermal current as $\CurrentEle$ and $\CurrentTh$, respectively. We assume these currents are caused by the external electric field and temperature gradient such as
\begin{align}
    \CurrentEle & = \LMat[11] \FieldEle + \LMat[12] \pqty{-\FieldTemperature},  \\
    \CurrentTh  & = \LMat[21] \FieldEle  + \LMat[22] \pqty{-\FieldTemperature},
\end{align}
where $\FieldEle$ is the external electric field, $\GradientTemperature$ is the temperature gradient, and $\LMat[ij]$ are the matrices of linear-response coefficients that depend on the material properties. We consider two directions, $x$ and $y$, and hence the vectors have two components and the matrices are $2 \times 2$. In the following, we assume isotropic materials, so that $\LMat[ij]$ is of the form
\begin{equation}
    \LMat[ij] = \mqty( \LDiag[ij] & \LNDiag[ij] \\ -\LNDiag[ij] & \LDiag[ij] ).
\end{equation}
The symmetry $\LMat[12] = \LMat[21]$ holds due to the Onsager reciprocal relations~\cite{Onsager1931}. While $\LMat[11]$ is the electric conductivity, the thermal conductivity $\CondThMat$, which is defined as the ratio of the thermal current to the temperature gradient at $\CurrentEle = \vb*{0}$, is given by
\begin{equation}
    \label{eq:intro_ThCondMat_def}
    \CondThMat = \frac{1}{\Temperature} \pqty{\LMat[22] - \LMat[21]\pqty{\LMat[11]}^{-1}\LMat[12]}.
\end{equation}

The Seebeck effect refers to occurrence of an electromotive force in a system with a temperature gradient~\cite{Seebeck1823, Seebeck1826}. To quantify the magnitude of this effect, the Seebeck coefficient ($\Seebeck$) is used, which is defined as the ratio of the electric field to the temperature gradient at $\CurrentEle = \vb*{0}$, $\Seebeck = \FieldEle[x] / \GradientTemperature[x]$. In the linear-response range, the Seebeck coefficient (without a magnetic field) can be expressed as
\begin{equation}
    \Seebeck = \frac{1}{\Temperature} \frac{\LDiag[12]}{\LDiag[11]},
\end{equation}
where we neglect the off-diagonal coefficients.

Apart from longitudinal thermoelectric effects, the transverse thermoelectric effects are characterized by the Nernst (Nernst-Ettingshausen) effect. The effect refers to the generation of a transverse electric field when a temperature gradient is applied to a material~\cite{Ettingshausen1886}. The Nernst coefficient $\Nernst$ is defined as the ratio of the transverse electric field to the temperature gradient under the condition of the absence of an electric current. Thus, considering the situation in which the temperature gradient is applied in the $x$ direction, the Nernst coefficient is expressed as
\begin{equation}
    \label{eq:def_ANC}
    \begin{split}
        \Nernst & = \frac{\FieldEle[y]}{\GradientTemperature[x]} = \frac{1}{\Temperature}\pqty{\pqty{\LMat[11]}^{-1}\LMat[12]}_{yx}   \\
                & = \frac{1}{T} \frac{\LNDiag[11]\LDiag[12] - \LDiag[11]\LNDiag[12]}{\pqty{\LDiag[11]}^{2} + \pqty{\LNDiag[11]}^{2}}.
    \end{split}
\end{equation}

In numerous instances, it is evident that there exists a straightforward equation that relates to $\LElem[ij]{\mu\nu}$ where $\mu, \nu \in \{x, y\}$. From the Boltzmann transport theory, one can derive the following relational equations satisfied by $\LElem[ij]{\mu\nu}$:
\begin{align}
    \label{eq:SB_L11}
    \LElem[11]{\mu\nu}                     & = \IntReg{-\infty}{\infty}{\ep}{
        \pqty{-\FermiDeriv[\ep]} \Sp{\mu\nu}[\ep, \Temperature]
    },                                                                                                 \\
    \label{eq:SB_L12}
    \LElem[12]{\mu\nu}= \LElem[21]{\mu\nu} & = \frac{1}{\EleCharge} \IntReg{-\infty}{\infty}{\ep}{
        \pqty{-\FermiDeriv[\ep]}\pqty{\ep - \Chemical}\Sp{\mu\nu}[\ep, \Temperature]
    },                                                                                                 \\
    \label{eq:SB_L22}
    \LElem[22]{\mu\nu}                     & = \frac{1}{\EleCharge^{2}} \IntReg{-\infty}{\infty}{\ep}{
        \pqty{-\FermiDeriv[\ep]} \pqty{\ep - \Chemical}^{2}\Sp{\mu\nu}[\ep, \Temperature]
    },
\end{align}
where $\EleCharge \pqty{ < 0}$ is the elementary charge,
$\Fermi(\ep)$ is the Fermi-Dirac distribution function defined as
\begin{equation}
    \Fermi(\ep) = \frac{1}{\exps{\InvKBT\pqty{\ep - \Chemical}} + 1},
\end{equation}
where $\InvKBT$ is the inverse temperature $1/\Boltzmann \Temperature$ and $\Chemical$ is the chemical potential with $\Boltzmann$ being the Boltzmann constant, and $\Sp{\mu\nu}[\ep, \Temperature]$ is a spectral function of conductivity.
We call the relational equations \eqref{eq:SB_L11}-\eqref{eq:SB_L22} the Sommerfeld-Bethe relation~\cite{Sommerfeld1933,Mott1936, Wilson1936}.

The Boltzmann equation has typically been  used to validate the accuracy of this relation~\cite{Mott1936}.
However, more recent research has provided a precise demonstration of its applicability using the microscopic Kubo linear-response theory for the $\LDiag[11]$ and $\LDiag[12]$~\cite{Jonson1980,Ogata2019}. It was shown that the Sommerfeld-Bethe relations hold in the presence of random-impurity potentials and periodic potentials. We extend the demonstration for $\LDiag[11]$ and $\LDiag[12]$ to $\LNDiag[11]$ and $\LNDiag[12]$, as detailed in Appendix~\ref{section:derivation_SB}. In the following, we consider the cases where the Sommerfeld-Bethe relation \eqref{eq:SB_L11}-\eqref{eq:SB_L22} hold for $\mu, \nu \in \{x, y\}$.

In metallic states, or more precisely, when $\Boltzmann \Temperature \ll |\mu-E_0|$ with $E_0$ being the band edge, the Sommerfeld expansion~\cite{Sommerfeld1928} leads to the well-known Mott formula~\cite{Mott1936} for the Seebeck coefficient
\begin{equation}
    \label{eq:Mott_formula}
    S = \frac{\LDiag[12]}{\Temperature\LDiag[11]} \simeq \frac{\pi^{2}}{3} \frac{\KBT}{\EleCharge} \left. \pdv{\pqty{\log \SpL}}{\ep} \right|_{\ep = \Chemical}.
\end{equation}
Similarly, for the ANC, the Sommerfeld expansion of Eq.~\eqref{eq:def_ANC} yields
\begin{equation}
    \label{eq:Mott_formula_ANC}
    \begin{split}
        \Nernst = & -\frac{\pi^{2}\Boltzmann^{2}\Temperature}{3\EleCharge}\frac{\SpT[\Chemical]/\SpL[\Chemical]}{1+\pqty{\SpT[\Chemical]/\SpL[\Chemical]}^{2}} \left. \dv{\ep} \log \frac{\SpT[\ep]}{\SpL[\ep]} \right|_{\ep = \Chemical} \\
                  & + \OrderLandau(\Temperature^{3}),
    \end{split}
\end{equation}
for $\Boltzmann \Temperature \ll |\mu-E_0|$.
Note that in the derivation of this formula, we have assumed that the spectral functions do not change significantly in the vicinity of the chemical potential and the spectral functions are non-zero at the chemical potential.
From Eq.~\eqref{eq:Mott_formula_ANC}, we can find that the ANC is completely determined by the behavior of the ratio $\SpT[\ep]/\SpL[\ep]$ at the chemical potential $\ep = \Chemical$. This formula is useful for understanding the behavior of the ANC in various systems. To obtain a large absolute value of the ANC in metals, the ratio $\SpT[\Chemical]/\SpL[\Chemical]$ should be near $1$, and the slope of the ratio should be steep at $\ep = \Chemical$.

\section{Mahan-Sofo-type Strategy for the Large Anomalous Nernst Coefficient}
\label{section:ANC_general}
In 1996, Mahan and Sofo~\cite{Mahan1996} recognized that the factor $-\FermiDeriv[\ep]\SpL[\ep, \Temperature]$ serves as a weight function in Eqs.~\eqref{eq:SB_L11}-\eqref{eq:SB_L22}. They then developed a strategy for achieving high longitudinal thermoelectric efficiency based on this insight. By defining a weighted average of $g(\ep)$ as
\begin{equation}
    \Mean{\mu\nu}{g} = \frac{\IntReg{-\infty}{\infty}{\ep}{\pqty{-\FermiDeriv[\ep]}\Sp{\mu\nu}[\ep, \Temperature] g(\ep)}}{\IntReg{-\infty}{\infty}{\ep}{\pqty{-\FermiDeriv[\ep]}\Sp{\mu\nu}[\ep, \Temperature]}},
\end{equation}

the Seebeck coefficient and the longitudinal thermal conductivity due to electrons can be expressed as
\begin{align}
    \Seebeck & = \frac{1}{\EleCharge T}\MeanL{\ep - \Chemical}, \\
    \begin{split}
        \CondTh_{\mathrm{ele}} & = \frac{1}{T}\pqty{\LDiag[22] - \frac{\LDiag[21]\LDiag[12]}{\LDiag[11]}}                                     \\
                               & = \frac{\LDiag[11]}{\EleCharge^{2}T}\pqty{\MeanL{\pqty{\ep - \Chemical}^{2}} - \MeanL{\ep - \Chemical}^{2}}.
    \end{split}
\end{align}
This expression indicates that the Seebeck coefficient is proportional to a kind of mean energy of the carriers, while the thermal conductivity is proportional to the variance of the energy. When the energy distribution becomes narrow, $\CondTh_{\mathrm{ele}}$ approaches zero. This reduction in electronic thermal conductivity is a key factor for achieving high $\zL \Temperature$ values.

Following the Mahan-Sofo approach, we develop a strategy for maximizing the ANC. The ANC expression in Eq.~\eqref{eq:def_ANC} can be rewritten as
\begin{align}
    \Nernst & = \frac{1}{\Temperature}\frac{\CondRatio}{1 + \CondRatio^{2}}
    \pqty{\frac{\LDiag[12]}{\LDiag[11]} - \frac{\LNDiag[12]}{\LNDiag[11]}} \cr
            & = \frac{1}{\EleCharge\Temperature}\frac{\CondRatio}{1+\CondRatio^{2}} \pqty{ \MeanL{\ep - \Chemical} - \MeanT{\ep - \Chemical}},
    \label{eq:ANC_spectre_expression}
\end{align}
with
\begin{equation}
    \label{eq:CondRatio}
    \CondRatio = \LNDiag[11]/\LDiag[11],
\end{equation}
which corresponds to the tangent of the Hall angle of the anomalous Hall effect.
Note that the transverse spectral function can be written as~\cite{Thouless1982}
\begin{equation}
    \SpT[\ep] = \frac{\EleCharge^{2}}{\hbar}\sum_{\vk} \Theta(\ep - \ep_{\vk}) \Omega_{\vk},
\end{equation}
where $\Theta(\ep)$ is the Heaviside step function, $\ep_{\vk}$ is the energy of the state with the momentum $\vk$, and $\Omega_{\vk}$ is the Berry curvature at the momentum $\vk$.

Equation \eqref{eq:ANC_spectre_expression} provides both a general framework and intuitive understanding of the ANC. Before we see examples in the next section, let us consider the general behavior of Eq.~\eqref{eq:ANC_spectre_expression}.

First, although $\CondRatio$ typically remains below $1$, a large value of $\CondRatio$ is essential for the ANC to be substantial. The prefactor $\CondRatio/(1+\CondRatio^2)$ is maximum at $\CondRatio = 1$.

Second, a significant disparity between the weight functions $-\FermiDeriv[\ep]\SpL[\ep]$ and $-\FermiDeriv[\ep]\SpT[\ep]$ is essential for large ANC values due to the factor  $\MeanL{\ep - \Chemical} - \MeanT{\ep - \Chemical}$. To maximize $\MeanL{\ep - \Chemical} - \MeanT{\ep - \Chemical}$, one must engineer spectral functions such that high-energy states preferentially contribute to longitudinal transport while low-energy states dominate transverse transport, or vice versa.

As in the Mahan-Sofo strategy, we look for the spectral functions that yield the maximum absolute value of ANC. The optimal configuration occurs when $\SpL(\ep)$ and $\SpT(\ep)$ exhibit delta-functional peaks, analogous to the Mahan-Sofo case~\cite{Mahan1996}. [Note: We assume that $\SpT(\ep)$ maintains a constant sign. Without this constraint, $|\MeanT{\ep - \Chemical}|$ could become arbitrarily large while keeping $\LNDiag[11]$ fixed.] Then, ANE becomes maximum when $\SpL(\ep)$ exhibits a delta-functional peak at some energy, say, $\ep = \Chemical +E$ while $\SpT(\ep)$ displays a similar peak at $\ep=\Chemical-E$, with the spectrum being zero in all other regions. In this case, $\CondRatio$ becomes $1$ and the ANC reaches $\abs{\Nernst} = \abs{E}/(\abs{\EleCharge}T)$. If we choose $\abs{E}=\Boltzmann T$, the value $\abs{\Nernst} = \Boltzmann/\abs{\EleCharge}$ is the same value as the absolute value of the Seebeck coefficient under the same situation.

This observation aligns with Mahan-Sofo's demonstration that $\zL T$ attains a high value when the spectral function exhibits a delta-function-like peak. Consequently, the aforementioned analysis provides a material search guideline for transverse thermoelectric response, corresponding to Mahan-Sofo's strategy.

\begin{figure}[tbp]
    \begin{center}
        \includegraphics[keepaspectratio, width=0.9\linewidth]{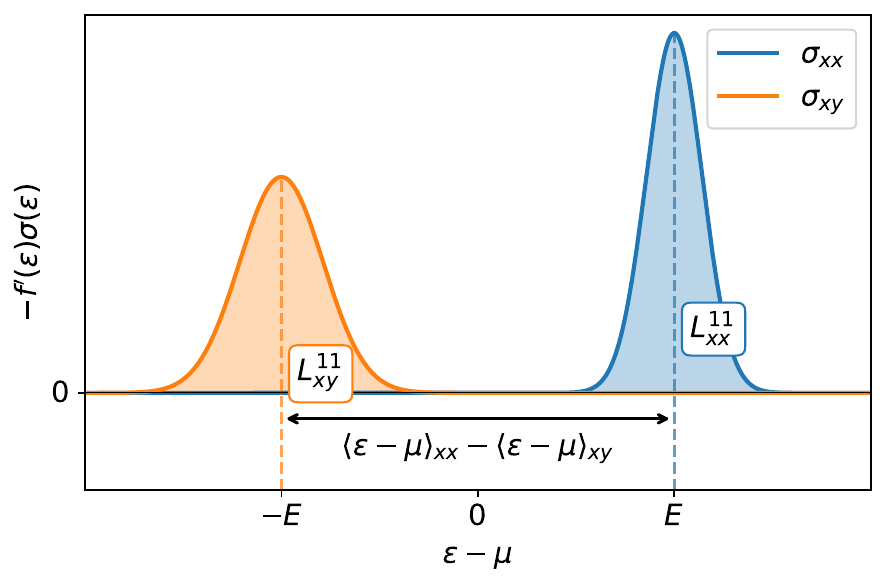}
        \caption{
            Conceptual illustration of the optimal spectral function configuration for maximizing the anomalous Nernst coefficient. The longitudinal spectral function $\SpL(\ep)$ (blue) and transverse spectral function $\SpT(\ep)$ (orange) peak at energies $\Chemical + E$ and $\Chemical - E$, respectively, where $\Chemical$ is the chemical potential. This separation results in the large anomalous Nernst coefficient according to our Mahan-Sofo-type strategy. The shaded areas under each spectral function correspond to the transport coefficients $\LDiag[11]$ and $\LNDiag[11]$, and the ANC becomes maximum when their ratio approaches unity ($\CondRatio = \LNDiag[11]/\LDiag[11] \rightarrow 1$).
        }
        \label{fig:strategy}
    \end{center}
\end{figure}

Figure~\ref{fig:strategy} illustrates our proposed strategy for achieving large ANC. The optimal configuration consists of longitudinal spectral function $\SpL(\ep)$ (blue) exhibiting a sharp peak at energy $\ep = \Chemical + E$ above the chemical potential, while the transverse spectral function $\SpT(\ep)$ (orange) displays a corresponding peak at $\ep = \Chemical - E$ below the chemical potential. This energy separation of $2E$ between the peaks generates the difference $\MeanL{\ep - \Chemical} - \MeanT{\ep - \Chemical}$ in Eq.~\eqref{eq:ANC_spectre_expression}, thereby optimizing the ANC. When $E = \Boltzmann T$, this configuration achieves the benchmark value $\abs{\Nernst} \simeq \Boltzmann/\abs{\EleCharge}$, which is comparable to the optimal Seebeck coefficient under similar conditions.

\section{Examples}
\label{section:examples}

In the subsequent analysis, we examine the ANC in three specific instances. Notably, the latter two examples serve as illustrative cases of spectral functions that can exhibit significant ANC.

\subsection{Ferromagnetic System}
First, let us see the ANC in ferromagnets discussed by Karplus and Luttinger~\cite{Karplus1954}.
According to their discussion and assumptions, the Hall conductivity is proportional to
\begin{equation}
    \sum_{\ell, \vk} \pqty{-\FermiDeriv[\ep_{\vk}]} \pqty{\Velocity[x]{\vk}[\ell]}^{2},
\end{equation}
where $\ell$ is the band index.
The longitudinal conductivity is proportional to the same factor.
Therefore, $\SpT[\ep]$ and $\SpL[\ep]$ have the same shape, and $\MeanL{\ep - \Chemical} - \MeanT{\ep - \Chemical}$ becomes zero, which leads the ANC to zero.
This is a simple explanation of the reason why ordinary ferromagnets do not show large ANCs. Of course, in real materials, the effects beyond their assumptions generate some ANCs, which, however, seem to be small considering the above discussion.

\begin{figure}[htbp]
    \begin{center}
        \includegraphics[keepaspectratio, width=0.9\linewidth]{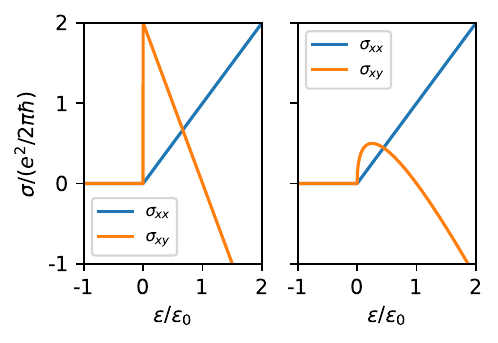}
        \caption{
            The illustrative spectral functions used in the example calculations of the ANC.
            The blue line represents the longitudinal spectral function, and the orange line represents the transverse spectral function.
            The left panel shows the linear spectral functions (see Eqs. \eqref{eq:SBN_linear_SpL} and \eqref{eq:SBN_linear_SpT}), and the right panel shows the square root spectral functions (see Eqs. \eqref{eq:SBN_root_SpL} and \eqref{eq:SBN_root_SpT}).
            In both cases, the coefficients are set to $A=1$, $B=-2$ and $r=-1$.
        }
        \label{fig:illustrative_spectrum}
    \end{center}
\end{figure}

\subsection{Linear Transverse Conductivity Spectrum}
Second, we consider the system with spectral functions
\begin{align}
    \SpL[\ep] & =
    \begin{dcases}
        \frac{\EleCharge^{2} A}{2\pi\hbar}  \frac{\ep}{\ep_{0}} & \pqty{\ep \geq 0}, \\
        0                                                       & \pqty{\ep < 0},
    \end{dcases}
    \label{eq:SBN_linear_SpL}
    \\
    \SpT[\ep] & =
    \begin{dcases}
        \frac{\EleCharge^{2} B}{2\pi\hbar}  \pqty{\frac{\ep}{\ep_{0}} + r} & \pqty{\ep \geq 0}, \\
        0                                                                  & \pqty{\ep < 0},
    \end{dcases}
    \label{eq:SBN_linear_SpT}
\end{align}
where $A$, $B$, and $r$ are dimensionless constants depending on the system and $\ep = 0$ is set to be the bottom of the band. $\ep_{0}$ is just a unit of energy, and it is positive. Typically, for example, in a system with Rashba interaction and Zeeman splitting, $\ep_{0}$ is about the order of the Zeeman splitting energy, and $r$ is about the order of 1.
The illustrative spectral functions are shown in the left panel of Fig.~\ref{fig:illustrative_spectrum}.
In this situation, the straightforward calculation leads to
\begin{align}
    \CondRatio & = \frac{\LNDiag[11]}{\LDiag[11]} = \frac{B}{A} \pqty{1 + \frac{r\InvKBT \ep_{0}}{\pqty{1 + \exps{-\InvKBT\Chemical}} \log \pqty{1 + \exps{\InvKBT\Chemical}}}},
    \label{eq:SBN_linear_Ratio}
    \\
    \Nernst    & = \frac{\Boltzmann}{\EleCharge}\frac{B}{A} \frac{r \InvKBT \ep_{0}}{1 + \CondRatio^{2}}
    \pqty{
        1-
        \frac{-2\PolyLog{2}{-\exps{\InvKBT\Chemical}}}
        {\pqty{\log\pqty{1+\exps{\InvKBT\Chemical}}}^{2}\pqty{1+\exps{-\InvKBT\Chemical}}}
    },
    \label{eq:SBN_linear_Nernst}
\end{align}
where $\InvKBT$ is the inverse temperature and $\PolyLog{s}{z}$ is the polylogarithm function defined as
\begin{equation}
    \PolyLog{s}{z} = \sum_{n=1}^{\infty} \frac{z^{n}}{n^{s}}.
\end{equation}
\par
\begin{figure}[tbp]
    \centering
    \subfigimg[width=0.85\columnwidth]{a}{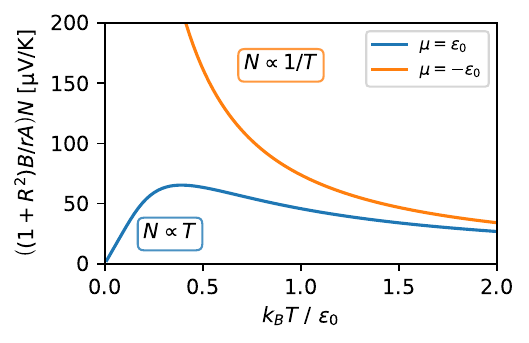}
    \centering
    \subfigimg[width=0.85\columnwidth]{b}{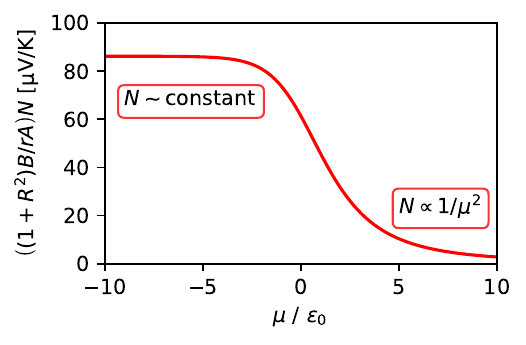}
    \caption{
        Temperature and chemical potential dependence of the anomalous Nernst coefficient from Eq.~\eqref{eq:SBN_linear_Nernst}.
        (a) the ANC as a function of temperature.
        We choose the unit of energy to be $\abs{\Chemical}$.
        We can see that when $\Chemical > 0$ (blue line) the ANC vanishes at low temperatures, while when $\Chemical < 0$ (orange line) the ANC diverges at low temperatures.
        (b) the ANC as a function of chemical potential at $\KBT = \ep_{0}$.
        We can see that the ANC converges to a finite value at low chemical potential. On the other hand, the ANC decreases as the chemical potential increases.
    }
    \label{fig:SBN_linear_Nernst}
\end{figure}

The ANC with these simple spectral functions is shown in Fig.~\ref{fig:SBN_linear_Nernst}. Figure~\ref{fig:SBN_linear_Nernst}(a) illustrates the ANC as a function of temperature. Here, we fix the chemical potential at $\Chemical=\ep_{0}$ or $\Chemical = -\ep_{0}$. When the chemical potential is negative, the ANC exhibits a divergence at low temperatures, roughly $\Nernst \sim 1/\Temperature$. Conversely, when the chemical potential is positive, the ANC diminishes at low temperatures, roughly $\Nernst \sim \Temperature$. This behavior is consistent with the Mott formula, Eq.~\eqref{eq:Mott_formula_ANC}. This suggests that the ANC can become substantial in low-doping metals or semiconductors. The difference between $\Chemical > 0$ and $\Chemical < 0$ becomes evident in Fig.~\ref{fig:SBN_linear_Nernst}(b), which shows the ANC as a function of chemical potential at $\KBT = \ep_{0}$. When the chemical potential is negatively large, the ANC converges to a finite value, $\Nernst \to (\Boltzmann/\EleCharge)(B/A)(r\InvKBT\ep_{0})/(1+\CondRatio^{2})$. In contrast, the ANC is diminished as the chemical potential increases, $\Nernst \sim 1/\Chemical^{2}$.

\subsection{Square Root Transverse Conductivity Spectrum}
Finally, we consider the system with spectral functions
\begin{align}
    \SpL[\ep] & =
    \begin{dcases}
        \frac{\EleCharge^{2} A}{2\pi\hbar}  \frac{\ep}{\ep_{0}} & \pqty{\ep \geq 0}, \\
        0                                                       & \pqty{\ep < 0},
    \end{dcases}
    \label{eq:SBN_root_SpL}
    \\
    \SpT[\ep] & =
    \begin{dcases}
        \frac{\EleCharge^{2} B}{2\pi\hbar}  \pqty{\frac{\ep}{\ep_{0}} + r \sqrt{\frac{\ep}{\ep_{0}}}} & \pqty{\ep \geq 0}, \\
        0                                                                                             & \pqty{\ep < 0}.
    \end{dcases}
    \label{eq:SBN_root_SpT}
\end{align}
The illustrative spectral functions are shown in the right panel of Fig.~\ref{fig:illustrative_spectrum}.
Following the same approach as in Eqs. \eqref{eq:SBN_linear_Ratio} and \eqref{eq:SBN_linear_Nernst}, we obtain
\begin{align}
    \CondRatio & = \frac{\LNDiag[11]}{\LDiag[11]} = \frac{B}{A} \pqty{1 + \frac{-r\sqrt{\pi\InvKBT\ep_{0}}\PolyLog{1/2}{-\exps{\InvKBT\Chemical}}}{2 \log \pqty{1 + \exps{\InvKBT\Chemical}}}}, \\
    \begin{split}
        \Nernst & = \frac{\Boltzmann}{\EleCharge}\frac{B}{A} \frac{r \sqrt{\pi\InvKBT \ep_{0}}}{1 + \CondRatio^{2}} \\
                & \times \pqty{
            \frac{-3\PolyLog{3/2}{-\exps{\InvKBT\Chemical}}}
            {4\log\pqty{1+\exps{\InvKBT\Chemical}}}
            - \frac{\PolyLog{2}{-\exps{\InvKBT\Chemical}}\PolyLog{1/2}{-\exps{\InvKBT\Chemical}}}{\pqty{\log\pqty{1+\exps{\InvKBT\Chemical}}}^{2}}
        }.
    \end{split}
    \label{eq:SBN_root_Nernst}
\end{align}

\begin{figure}[tbp]
    \centering
    \subfigimg[width=0.85\columnwidth]{a}{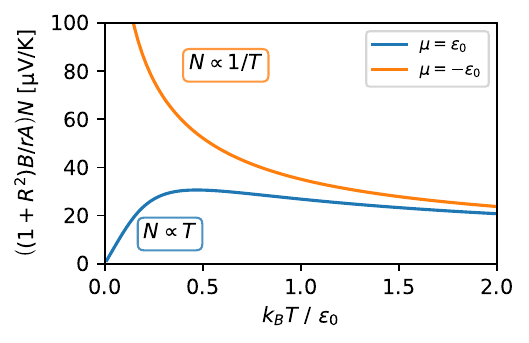}
    \centering
    \subfigimg[width=0.85\columnwidth]{b}{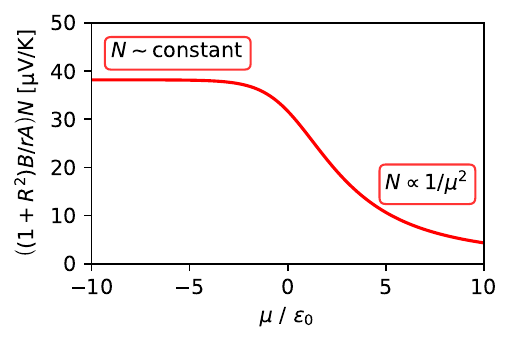}
    \caption{
        Temperature and chemical potential dependence of the anomalous Nernst coefficient from Eq.~\eqref{eq:SBN_root_Nernst}.
        (a) the ANC as a function of temperature.
        We choose the unit of energy to be $\abs{\Chemical}$.
        We can see that when $\Chemical > 0$ (blue line) the ANC vanishes at low temperatures, while when $\Chemical < 0$ (orange line) the ANC diverges at low temperatures.
        (b) the ANC as a function of chemical potential at $\KBT = \ep_{0}$.
        We can see that the ANC converges to finite value at low chemical potential.
        On the other hand, the ANC decreases as the chemical potential increases.
    }
    \label{fig:SBN_root_Nernst}
\end{figure}

The ANC with this square-root spectral functions is shown in Fig.~\ref{fig:SBN_root_Nernst}. Figure~\ref{fig:SBN_root_Nernst}(a) illustrates the ANC as a function of temperature. In contrast to Fig.~\ref{fig:SBN_linear_Nernst}(a), the ANC exhibits a divergence at low temperatures, roughly as $\Nernst \sim 1/\sqrt{\Temperature}$ when the chemical potential is negative.
Figure \ref{fig:SBN_root_Nernst}(b) shows the ANC as a function of chemical potential at $\KBT = \ep_{0}$. At low chemical potential, the ANC converges to a finite value as $\Nernst \to (\sqrt{\pi}/4)(\Boltzmann/\EleCharge)(B/A)(r\sqrt{\InvKBT\ep_{0}})/(1+\CondRatio^{2})$. In contrast, the ANC diminishes as the chemical potential increases, $N \sim 1/\Chemical^{2}$.

To summarize, there must be a difference between $\SpL[\ep]$ and $\SpT[\ep]$ in order for the ANC to be non-zero. In a simple ferromagnetic model, there is either a negligible or no ANC. However, dilute metals or semiconductors with spectral functions represented by Eqs.~\eqref{eq:SBN_linear_SpT} or \eqref{eq:SBN_root_SpT} can exhibit a significant magnitude of the ANC.

\section{Conclusion}
\label{section:conclusion}
We proposed a strategy for enhancing the anomalous Nernst coefficient based on the Sommerfeld-Bethe relation. Our systematic analysis establishes the conditions necessary for significant enhancement of the anomalous Nernst coefficient. Using a general expression Eq.~\eqref{eq:ANC_spectre_expression}, we found that the difference between the longitudinal and transverse spectral functions is crucial for achieving a large anomalous Nernst coefficient. At low temperatures, the coefficient magnitude is determined by the behavior of the ratio of transverse to longitudinal spectral functions at the chemical potential, as demonstrated in Eq.~\eqref{eq:Mott_formula_ANC}.

We also introduced simplified models that demonstrate enhanced coefficients. Although simple ferromagnetic models cannot achieve a large anomalous Nernst coefficient, we found that a model with a transverse spectral function that exhibits a singular behavior at the band edge can yield a large anomalous Nernst coefficient. This model provides a guideline for designing materials with large anomalous Nernst coefficients.

The next step is the concrete calculation based on the realistic model. The system with Rashba interaction and Zeeman effect is a good candidate for understanding the strategy suggested in this paper. This will be discussed elsewhere.

\begin{acknowledgments}
    J. E. was supported by the Japan Society for the Promotion of Science and the Leading House ETH Zurich through the Program for Leading Graduate Schools (MERIT) and Young Researchers' Exchange Programme (No. JP\_EG\_special\_032023\_16).
This work is supported by Grants-in-Aid for Scientific Research from the Japan Society for the Promotion of Science (Nos. JP22KJ1055, JP22KK0228, and JP23K03274), and JST-Mirai Program Grant (No. JPMJMI19A1).
\end{acknowledgments}

\appendix
\begin{widetext}
    \section{Derivation of the Sommerfeld-Bethe relation}
    \label{section:derivation_SB}
    In this appendix, we derive the Sommerfeld-Bethe relation between $\LDiag[11]$ and $\LDiag[12]$, and between $\LNDiag[11]$ and $\LNDiag[12]$. The derivation is based on Refs.~\cite{Ogata2019} and \cite{Jonson1990}. In Section \ref{sec:Appendix_derivation_SB_main}, we derive the Sommerfeld-Bethe relation under a specific condition for the thermal current operators. In Section \ref{sec:Appendix_derivation_SB_thermal_current}, we examine the condition of the thermal current operators for several Hamiltonians.

\subsection{Main Part of the Derivation}
\label{sec:Appendix_derivation_SB_main}

Let us consider a model with a Hamiltonian $\Hamiltonian$. The electric current operator is assumed to be given by
\begin{equation}
    \OpCurrentEle = \sum_{\vk} \OpCD[\vk] \CurrentDensEle[\vk] \OpC[\vk],
\end{equation}
where $\CurrentDensEle[\vk]$ is the electric current density operator for the electron with momentum $\vk$.
Here, $\OpC$ and $\OpCD$ are the annihilation and creation operators of the electron with momentum $\vk$, respectively, which can contain the spin index, the band index, and some other indices.
In addition, we define the imaginary-time representation of the operators as
\begin{align}
    \OpC(\tau)  & = \exps{\tau \Hamiltonian} \OpC \exps{-\tau \Hamiltonian},  \\
    \OpCD(\tau) & = \exps{\tau \Hamiltonian} \OpCD \exps{-\tau \Hamiltonian},
\end{align}
which satisfy the equations of motion given by
\begin{align}
    \pdv{\tau} \OpC(\tau)  & = \comm{\Hamiltonian}{\OpC(\tau)},  \\
    \pdv{\tau} \OpCD(\tau) & = \comm{\Hamiltonian}{\OpCD(\tau)}.
\end{align}
Using them, the imaginary-time representation of the electric current operator is given by
\begin{equation}
    \OpCurrentEle(\tau) = \lim_{\tau' \nearrow \tau} \sum_{\vk} \OpCD[\vk](\tau) \CurrentDensEle[\vk] \OpC[\vk](\tau').
\end{equation}
In this section, we assume that the thermal current operator is given by
\begin{equation}
    \label{eq:Appendix_derivation_SB_thermal_current_assumption}
    \OpCurrentTh = \frac{1}{2\EleCharge}\sum_{\vk} \pqty{
        \comm{\Hamiltonian - \Chemical \OpNum}{\OpCD[\vk]} \CurrentDensEle[\vk] \OpC[\vk] - \OpCD[\vk] \CurrentDensEle[\vk] \comm{\Hamiltonian - \Chemical \OpNum}{\OpC[\vk]}
    },
\end{equation}
where $\Chemical$ is the chemical potential and $\OpNum$ is the number operator.

We define a function $F(\tau , \tau')$ as
\begin{equation}
    F(\tau , \tau') = \mean{\sum_{\vk} \OpCD[\vk](\tau) \CurrentDensEle[\vk] \OpC[\vk](\tau') \OpCurrentEle(0)},
\end{equation}
where $\tau$ and $\tau'$ are assumed to be positive. It satisfies
\begin{equation}
    \lim_{\tau' \nearrow \tau}F(\tau , \tau') = \mean{\TauOrder[\OpCurrentEle(\tau) \OpCurrentEle(0)]}.
\end{equation}
Considering the assumption Eq.~\eqref{eq:Appendix_derivation_SB_thermal_current_assumption} and the equations of motion, we obtain
\begin{equation}
    \lim_{\tau' \nearrow \tau}\frac{1}{2\EleCharge}\pqty{\pdv{\tau} - \pdv{\tau'}} F(\tau , \tau') = \mean{\TauOrder[\OpCurrentTh(\tau) \OpCurrentEle(0)]}.
\end{equation}

Let us calculate the correlation functions. Using the Bloch-de Dominicis theorem, we obtain
\begin{equation}
    \begin{split}
         & \quad \Correlation[11][\mu\nu][\vb*{0}][\iwlam]                                                                    \\
         & = \frac{1}{\Volume} \IntReg{0}{\InvKBT}{\tau}{ \pqty{\lim_{\tau' \nearrow \tau} F(\tau,\tau')} \exps{\iwlam \tau}} \\
         & = -\frac{1}{\Volume} \sum_{\ell, \ell', \ell'', \ell'''}\IntReg{0}{\InvKBT}{\tau}{
            \pqty{\CurrentDensEle[\vk,\mu]}_{\ell\ell'}\mean{\TauOrder[\OpC[\ell'](\tau) \OpCD[\ell''](0)]}
            \pqty{\CurrentDensEle[\vk,\nu]}_{\ell''\ell'''}\mean{\TauOrder[\OpC[\ell'''](0)\OpCD[\ell](\tau)]} \exps{\iwlam \tau}
        }                                                                                                                     \\
         & = - \sum_{n,\vk} \Tr \bqty{
            \CurrentDensEle[\vk,\mu] \GreT[\vk][\ien] \CurrentDensEle[\vk,\nu]\GreT[\vk][\ien - \iwlam]
        }.
    \end{split}
\end{equation}
Taking the summation of the Matsubara frequency and the analytic continuation $\iwlam \to \hbar\pqty{\omega + i \delta}$, we obtain
\begin{equation}
    \begin{split}
         & \quad \Correlation[11][\mu\nu][\vb*{0}][\hbar\pqty{\omega + i \delta}]                                                  \\
         & = \frac{1}{2\pi i} \sum_{\vk} \IntReg{-\infty}{\infty}{\ep}{
            \Fermi(\ep) \Tr \left[
        \CurrentDensEle[\vk,\mu] \GreR[\vk][\ep + \hbar\omega] \CurrentDensEle[\vk,\nu] \GreR[\vk][\ep] \right.                    \\
         & \qquad - \CurrentDensEle[\vk,\mu] \GreR[\vk][\ep + \hbar\omega] \CurrentDensEle[\vk,\nu] \GreA[\vk][\ep]                \\
         & \qquad + \CurrentDensEle[\vk,\mu] \GreR[\vk][\ep] \CurrentDensEle[\vk,\nu] \GreA[\vk][\ep - \hbar\omega]                \\
         & \left. \qquad - \CurrentDensEle[\vk,\mu] \GreA[\vk][\ep] \CurrentDensEle[\vk,\nu] \GreA[\vk][\ep - \hbar\omega] \right]
        }.
    \end{split}
\end{equation}
Therefore, the conductivity is given by
\begin{equation}
    \label{eq:Appendix_derivation_SB_main_L11}
    \begin{split}
        L^{11}_{\mu\nu} & = \frac{\Correlation[11][\mu\nu][\vb*{0}][\omega + i \delta] - \Correlation[11][\mu\nu][\vb*{0}][0]}{\hbar\pqty{\omega + i \delta}} \\
                        & = \frac{1}{2\pi}\sum_{\vk}\IntReg{-\infty}{\infty}{\ep}{\pqty{-\Fermi'(\ep)}
            \Tr\left[\CurrentDensEle[\vk,\mu]\GreR[\vk][\ep]\CurrentDensEle[\vk,\nu]\GreA[\vk][\ep]
        \right.                                                                                                                                               \\
                        & \left. \quad - \frac{1}{2}\pqty{
                    \CurrentDensEle[\vk,\mu]\GreR[\vk][\ep]\CurrentDensEle[\vk,\nu]\GreR[\vk][\ep]
                    +\CurrentDensEle[\vk,\mu]\GreA[\vk][\ep]\CurrentDensEle[\vk,\nu]\GreA[\vk][\ep]
                }
                \right]
        }.
    \end{split}
\end{equation}

On the other hand, we can calculate the correlation function $\Correlation[12][\mu\nu][\vb*{0}][\iwlam]$ as
\begin{equation}
    \begin{split}
         & \quad \Correlation[21][\mu\nu][\vb*{0}][\iwlam]                                                                                                                   \\
         & = \frac{1}{V} \IntReg{0}{\InvKBT}{\tau}{ \pqty{\lim_{\tau' \nearrow \tau} \frac{1}{2\EleCharge}\pqty{\pdv{\tau} - \pdv{\tau'}} F(\tau,\tau')} \exps{\iwlam \tau}} \\
         & = -\frac{1}{\Volume} \sum_{\ell, \ell', \ell'', \ell'''}\IntReg{0}{\InvKBT}{\tau}{
        \exps{\iwlam \tau}\pqty{\CurrentDensEle[\vk,\mu]}_{\ell\ell'}\pqty{\CurrentDensEle[\vk,\nu]}_{\ell''\ell'''}                                                         \\
         & \quad   \times \frac{1}{2\EleCharge}
            \left\{
            \mean{\TauOrder[\OpC[\ell'](\tau) \OpCD[\ell''](0)]}
            \pqty{\pdv{\tau}\mean{\TauOrder[\OpC[\ell'''](0)\OpCD[\ell](\tau)]}}
        \right.                                                                                                                                                              \\
         & \qquad \left.
            - \pqty{\pdv{\tau}\mean{\TauOrder[\OpC[\ell'](\tau) \OpCD[\ell''](0)]}}
            \mean{\TauOrder[\OpC[\ell'''](0)\OpCD[\ell](\tau)]}
            \right\}
        }                                                                                                                                                                    \\
         & = - \frac{1}{\EleCharge}\sum_{n,\vk} \pqty{\ien - \frac{\iwlam}{2}} \Tr \bqty{
            \CurrentDensEle[\vk,\mu] \GreT[\vk][\ien] \CurrentDensEle[\vk,\nu]\GreT[\vk][\ien - \iwlam]
        }.
    \end{split}
\end{equation}
In the same manner as the previous calculation, we obtain
\begin{equation}
    \begin{split}
         & \quad \Correlation[21][\mu\nu][\vb*{0}][\hbar\pqty{\omega + i \delta}]                                                                                           \\
         & = \frac{1}{2\pi i \EleCharge} \sum_{\vk} \IntReg{-\infty}{\infty}{\ep}{
            \Fermi(\ep) \Tr \left[
        \pqty{\ep - \mu + \frac{\hbar\omega}{2}}\CurrentDensEle[\vk,\mu] \GreR[\vk][\ep + \hbar\omega] \CurrentDensEle[\vk,\nu] \GreR[\vk][\ep] \right.                     \\
         & \qquad - \pqty{\ep - \mu + \frac{\hbar\omega}{2}} \CurrentDensEle[\vk,\mu] \GreR[\vk][\ep + \hbar\omega] \CurrentDensEle[\vk,\nu] \GreA[\vk][\ep]                \\
         & \qquad + \pqty{\ep - \mu - \frac{\hbar\omega}{2}} \CurrentDensEle[\vk,\mu] \GreR[\vk][\ep] \CurrentDensEle[\vk,\nu] \GreA[\vk][\ep - \hbar\omega]                \\
         & \left. \qquad - \pqty{\ep - \mu - \frac{\hbar\omega}{2}} \CurrentDensEle[\vk,\mu] \GreA[\vk][\ep] \CurrentDensEle[\vk,\nu] \GreA[\vk][\ep - \hbar\omega] \right]
        } ,
    \end{split}
\end{equation}
and the conductivity is given by
\begin{equation}
    \label{eq:Appendix_derivation_SB_main_L21}
    \begin{split}
        L^{21}_{\mu\nu} & = \frac{1}{\EleCharge}\frac{\Correlation[12][\mu\nu][\vb*{0}][\omega + i \delta] - \Correlation[12][\mu\nu][\vb*{0}][0]}{\omega + i \delta} \\
                        & = \frac{1}{2\pi\EleCharge}\sum_{\vk}\IntReg{-\infty}{\infty}{\ep}{\pqty{-\Fermi'(\ep)}\pqty{\ep - \mu}
            \Tr\left[\CurrentDensEle[\vk,\mu]\GreR[\vk][\ep]\CurrentDensEle[\vk,\nu]\GreA[\vk][\ep]
        \right.                                                                                                                                                       \\
                        & \left. \quad - \frac{1}{2}\pqty{
                    \CurrentDensEle[\vk,\mu]\GreR[\vk][\ep]\CurrentDensEle[\vk,\nu]\GreR[\vk][\ep]
                    +\CurrentDensEle[\vk,\mu]\GreA[\vk][\ep]\CurrentDensEle[\vk,\nu]\GreA[\vk][\ep]
                }
                \right]
        }.
    \end{split}
\end{equation}
Comparing Eqs.~\eqref{eq:Appendix_derivation_SB_main_L11} and \eqref{eq:Appendix_derivation_SB_main_L21}, we obtain the Sommerfeld-Bethe relation. For $\mu = \nu = x$, we obtain the relation between $\LDiag[11]$ and $\LDiag[12]$, whereas for $\mu = x$ and $\nu = y$, we obtain the relation between $\LNDiag[11]$ and $\LNDiag[12]$.

It is important to note that the relation can incorporate vertex corrections, which can be derived in a straightforward manner. Then, one current operator is substituted with the vertex-corrected current operator.

\subsection{Several Thermal Current Operators}
\label{sec:Appendix_derivation_SB_thermal_current}

In Section \ref{sec:Appendix_derivation_SB_main}, we assumed the relation between the electric current operator and the thermal current operator as expressed in Eq.~\eqref{eq:Appendix_derivation_SB_thermal_current_assumption}. The derivation demonstrates that this relation constitutes a necessary and sufficient condition for the Sommerfeld-Bethe relation. This section examines the aforementioned condition for various Hamiltonians.

Throughout this section, we derive the current operators by the equation of continuity given by
\begin{align}
    \label{eq:Appendix_derivation_SB_JEle_continuity_real}
    \Div \OpCurrentEle (\vrr)  & = \frac{i\EleCharge}{\hbar}\comm{\OpDens(\vrr)}{\Hamiltonian},   \\
    \label{eq:Appendix_derivation_SB_JTh_continuity_real}
    \Div \OpCurrentHeat (\vrr) & = \frac{i}{\hbar}\comm{\HamiltonianDensity(\vrr)}{\Hamiltonian},
\end{align}
where $\OpDens$ is the charge density operator, and $\HamiltonianDensity$ is the Hamiltonian density operator. The operator $\OpCurrentHeat$ is the heat current operator, which satisfies
\begin{equation}
    \OpCurrentTh = \OpCurrentHeat - \frac{\Chemical}{\EleCharge}\OpCurrentEle.
\end{equation}
In the wave-number representation, the continuity equations are given by
\begin{align}
    \label{eq:Appendix_derivation_SB_JEle_continuity_wave}
    \vq \cdot \OpCurrentEle_{\vq}  & = \frac{\EleCharge}{\hbar}\comm{\OpDens_{\vq}}{\Hamiltonian},    \\
    \label{eq:Appendix_derivation_SB_JTh_continuity_wave}
    \vq \cdot \OpCurrentHeat_{\vq} & = \frac{1}{\hbar}\comm{\HamiltonianDensity_{\vq}}{\Hamiltonian},
\end{align}
where the Fourier transformation of operator $\hat{A}(\vrr)$ is defined by
\begin{equation}
    \hat{A}_{\vq} = \Int{\vrr}{\exps{-i\vq \cdot \vrr}\hat{A}(\vrr)}.
\end{equation}

Using these equations, we examine the relation
\begin{equation}
    \label{eq:Appendix_derivation_SB_heat_current_assumption}
    \OpCurrentHeat = \frac{1}{2\EleCharge} \sum_{\vk} \pqty{
        \comm{\Hamiltonian}{\OpCD[\vk]} \CurrentDensEle[\vk] \OpC[\vk] - \OpCD[\vk] \CurrentDensEle[\vk] \comm{\Hamiltonian}{\OpC[\vk]}
    },
\end{equation}
which is equivalent to Eq.~\eqref{eq:Appendix_derivation_SB_thermal_current_assumption}.

\subsubsection{Free electrons}
\label{subsec:Appendix_derivation_SB_thermal_current_free_electron}

For free electrons, the Hamiltonian is given by
\begin{equation}
    \Hamiltonian = \sum_{\vk} \OpCD[\vk]\ep_{\vk} \OpC[\vk],
\end{equation}
where $\ep_{\vk} =  \hbar^2 \abs{\vk}^2/2\MassEle$.
The density operator is given by
\begin{equation}
    \OpDens_{\vq} = \sum_{\vk} \OpCD[\vk - \vq] \OpC[\vk].
\end{equation}
From Eq.~\eqref{eq:Appendix_derivation_SB_JEle_continuity_wave}, we obtain
\begin{equation}
    \begin{split}
        \vq \cdot \OpCurrentEle_{\vq}
         & = \frac{\EleCharge}{\hbar}\comm{\OpDens_{\vq}}{\Hamiltonian}                                                                                                 \\
         & = \frac{\EleCharge}{\hbar}\comm{\sum_{\vk}\OpCD[\vk - \vq] \OpC[\vk]}{\sum_{\vk'} \OpCD[\vk'] \frac{\hbar^2 \abs{\vk'}^2}{2\MassEle} \OpC[\vk']}             \\
         & = \frac{\EleCharge}{\hbar} \sum_{\vk,\vk'} \frac{\hbar^2 \abs{\vk'}^2}{2\MassEle}\pqty{
            \OpCD[\vk - \vq] \acomm{\OpC[\vk]}{\OpCD[\vk']} \OpC[\vk'] - \OpCD[\vk'] \acomm{\OpC[\vk-\vq]}{\OpCD[\vk']}\OpC[\vk]
        }                                                                                                                                                               \\
         & = \frac{\EleCharge}{\hbar} \sum_{\vk}  \OpCD[\vk - \vq] \pqty{\frac{\hbar^2 \abs{\vk}^2}{2\MassEle} - \frac{\hbar^2 \abs{\vk - \vq}^2}{2\MassEle}} \OpC[\vk] \\
         & \simeq \vq\cdot \EleCharge \sum_{\vk}  \OpCD[\vk - \vq] \Velocity{\vk} \OpC[\vk],
    \end{split}
\end{equation}
where $ \Velocity{\vk} = \hbar \vk/\MassEle$, which leads to $\OpCurrentEle_{\vq} = \EleCharge \sum_{\vk}  \OpCD[\vk- \vq] \Velocity{\vk} \OpC[\vk]$. It should be noted that, in a strict sense, the velocity operator should be $ \Velocity{\vk - \vq / 2}$ instead of $ \Velocity{\vk}$. However, the difference is negligible as we approach the limit $\vq \to 0$. In subsequent discussions, similar approximations are employed.

Let us turn to the heat current operator. The Hamiltonian density operator is given by
\begin{equation}
    \HamiltonianDensity_{\vq} = \sum_{\vk} \OpCD[\vk - \vq] \ep_{\vk} \OpC[\vk].
\end{equation}
From Eq.~\eqref{eq:Appendix_derivation_SB_JTh_continuity_wave}, we obtain
\begin{equation}
    \begin{split}
        \vq \cdot \OpCurrentHeat_{\vq}
         & = \frac{1}{\hbar}\comm{\HamiltonianDensity_{\vq}}{\Hamiltonian}                                                                                                               \\
         & = \frac{1}{\hbar}\comm{\sum_{\vk}\OpCD[\vk - \vq] \frac{\hbar^2 \abs{\vk}^2}{2\MassEle} \OpC[\vk]}{\sum_{\vk'} \OpCD[\vk'] \frac{\hbar^2 \abs{\vk'}^2}{2\MassEle} \OpC[\vk']} \\
         & =  \frac{1}{\hbar} \sum_{\vk,\vk'} \frac{\hbar^2 \abs{\vk}^2}{2\MassEle}\frac{\hbar^2 \abs{\vk'}^2}{2\MassEle}\pqty{
            \OpCD[\vk - \vq] \acomm{\OpC[\vk]}{\OpCD[\vk']} \OpC[\vk'] - \OpCD[\vk'] \acomm{\OpC[\vk-\vq]}{\OpCD[\vk']}\OpC[\vk]
        }                                                                                                                                                                                \\
         & \simeq \vq\cdot \sum_{\vk}  \OpCD[\vk - \vq]  \Velocity{\vk} \ep_{\vk} \OpC[\vk],
    \end{split}
\end{equation}
which leads to $\OpCurrentHeat_{\vq} = \sum_{\vk}  \OpCD[\vk- \vq] \Velocity{\vk}  \ep_{\vk} \OpC[\vk]$. On the other hand, the commutation relations read
\begin{align}
    \begin{split}
        \comm{\Hamiltonian}{\OpCD[\vk]} & = \comm{\sum_{\vk'} \OpCD[\vk'] \ep_{\vk'} \OpC[\vk']}{\OpCD[\vk]} \\
                                        & = \sum_{\vk'} \ep_{\vk'} \OpCD[\vk']\acomm{\OpC[\vk']}{\OpCD[\vk]} \\
                                        & = \ep_{\vk} \OpCD[\vk],
    \end{split}
    \\
    \begin{split}
        \comm{\Hamiltonian}{\OpC[\vk]} & = \comm{\sum_{\vk'} \OpCD[\vk'] \ep_{\vk'} \OpC[\vk']}{\OpC[\vk]}  \\
                                       & = -\sum_{\vk'} \ep_{\vk'}\acomm{\OpCD[\vk']}{\OpC[\vk]} \OpC[\vk'] \\
                                       & = -\ep_{\vk} \OpC[\vk].
    \end{split}
\end{align}
Therefore, we obtain the relation Eq.~\eqref{eq:Appendix_derivation_SB_heat_current_assumption} for $\OpCurrentHeat = \OpCurrentHeat_{\vb*{0}}$.

\subsubsection{Bloch Electrons}
\label{subsec:Appendix_derivation_SB_thermal_current_bloch_electron}

Next, we consider the Hamiltonian for Bloch electrons under a periodic potential, which is given by
\begin{equation}
    \Hamiltonian = \sum_{\vk, \ell} \OpCD[\ell,\vk] \ep_{\ell,\vk} \OpC[\ell,\vk],
\end{equation}
where $\ep_{\ell,\vk}$ is the energy dispersion of the $\ell$th band. The density operator is given by
\begin{equation}
    \label{eq:Appendix_derivation_SB_Bloch_density_operator}
    \OpDens_{\vq} = \sum_{\vk, \ell, \ell'} \OpCD[\ell',\vk - \vq] \braket{\Bloch{\vk-\vq}[\ell']}{\Bloch{\vk}[\ell]} \OpC[\ell,\vk],
\end{equation}
where $\braket{\Bloch{\vk-\vq}[\ell']}{\Bloch{\vk}[\ell]}$ is the inner product of the Bloch wave functions defined by
\begin{equation}
    \braket{\Bloch{\vk-\vq}[\ell']}{\Bloch{\vk}[\ell]} = \Int{\vrr}{\Bloch{\vk-\vq}[\ell]^{*}(\vrr) \Bloch{\vk}[\ell](\vrr)}.
\end{equation}
Note that $\braket{\Bloch{\vk}[\ell']}{\Bloch{\vk}[\ell]} = \delta_{\ell',\ell}$ holds.
The electric current operator is obtained by
\begin{equation}
    \begin{split}
        \vq \cdot \OpCurrentEle_{\vq}
         & = \frac{\EleCharge}{\hbar}\comm{\OpDens_{\vq}}{\Hamiltonian}                                                                                                                                                             \\
         & = \frac{\EleCharge}{\hbar}\comm{\sum_{\vk',\ell',\ell''}\OpCD[\ell'',\vk' - \vq] \braket{\Bloch{\vk'-\vq}[\ell'']}{\Bloch{\vk'}[\ell']} \OpC[\ell',\vk']}{\sum_{\vk,\ell} \OpCD[\ell,\vk] \ep_{\ell,\vk} \OpC[\ell,\vk]} \\
         & = \frac{\EleCharge}{\hbar} \sum_{\vk,\ell,\ell'} \OpCD[\ell',\vk - \vq] \braket{\Bloch{\vk-\vq}[\ell']}{\Bloch{\vk}[\ell]} \pqty{\ep_{\ell,\vk} - \ep_{\ell',\vk - \vq}} \OpC[\ell,\vk]                                  \\
         & \simeq \vq \cdot \EleCharge \pqty{
            \sum_{\vk, \ell} \OpCD[\ell,\vk - \vq] \Velocity{\vk}[\ell] \OpC[\ell,\vk]
            + \frac{1}{\hbar}\sum_{\vk, \ell \neq \ell'} \OpCD[\ell',\vk - \vq] \mel{\Bloch{\vk - \vq}[\ell']}{\pdv{\vk}}{\Bloch{\vk}[\ell]}\pqty{\ep_{\ell,\vk} - \ep_{\ell',\vk - \vq}} \OpC[\ell,\vk]
        },
    \end{split}
\end{equation}
where $\Velocity{\vk}[\ell]$ is the velocity operator for the $\ell$th band given by $\partial \ep_{\ell,\vk} / \hbar \partial \vk$. The first term is the intra-band current operator and the second term is the inter-band current operator.

Since the Hamiltonian density operator is given by
\begin{equation}
    \HamiltonianDensity_{\vq} = \frac{1}{2}\sum_{\vk,\ell} \OpCD[\ell', \vk - \vq]  \braket{\Bloch{\vk-\vq}[\ell']}{\Bloch{\vk}[\ell]} \pqty{\ep_{\ell,\vk} + \ep_{\ell', \vk - \vq}} \OpC[\ell,\vk],
\end{equation}
the heat current operator is derived from
\begin{equation}
    \begin{split}
        \vq \cdot \OpCurrentHeat_{\vq}
         & = \frac{1}{2\hbar}\comm{\HamiltonianDensity_{\vq}}{\Hamiltonian}                                                                                                                                                                                                \\
         & = \frac{1}{2\hbar}\comm{\sum_{\vk',\ell',\ell''}\OpCD[\ell'',\vk' - \vq] \braket{\Bloch{\vk'-\vq}[\ell'']}{\Bloch{\vk'}[\ell']} \pqty{\ep_{\ell'',\vk-\vq} + \ep_{\ell', \vk}} \OpC[\ell',\vk']}{\sum_{\vk,\ell} \OpCD[\ell,\vk] \ep_{\ell,\vk} \OpC[\ell,\vk]} \\
         & = \frac{1}{2\hbar} \sum_{\vk,\ell,\ell'} \OpCD[\ell',\vk - \vq] \braket{\Bloch{\vk-\vq}[\ell']}{\Bloch{\vk}[\ell]} \pqty{\ep_{\ell,\vk} - \ep_{\ell',\vk - \vq}}  \pqty{\ep_{\ell,\vk} + \ep_{\ell',\vk - \vq} } \OpC[\ell,\vk]                                 \\
         & \simeq \vq \cdot \left(
        \sum_{\vk, \ell} \OpCD[\ell,\vk - \vq] \Velocity{\vk}[\ell]\ep_{\ell,\vk} \OpC[\ell,\vk] \right.                                                                                                                                                                   \\
         & \quad  \left. + \frac{1}{2\hbar}\sum_{\vk, \ell \neq \ell'} \OpCD[\ell',\vk - \vq]  \mel{\Bloch{\vk-\vq}[\ell']}{\pdv{\vk}}{\Bloch{\vk}[\ell]}\pqty{\ep_{\ell,\vk} - \ep_{\ell',\vk - \vq}}\pqty{\ep_{\ell,\vk} + \ep_{\ell',\vk - \vq}}\OpC[\ell,\vk]
        \right).
    \end{split}
\end{equation}
Additionally, the following commutation relations hold:
\begin{align}
    \comm{\Hamiltonian}{\OpCD[\ell,\vk]}
     & = \ep_{\ell,\vk} \OpCD[\ell,\vk],
    \\
    \comm{\Hamiltonian}{\OpC[\ell,\vk]}
     & = -\ep_{\ell,\vk} \OpC[\ell,\vk].
\end{align}
Therefore, the relation Eq.~\eqref{eq:Appendix_derivation_SB_heat_current_assumption} obviously holds.

\end{widetext}

\bibliography{reference}

\end{document}